\documentclass[twocolumn,pre,showpacs]{revtex4}
\usepackage{amssymb, amsmath, graphicx}
\def\be{\begin{equation}}
\def\ee{\end{equation}}
\begin{document}
\title{Spectral Density Scaling of Fluctuating Interfaces}
\author{Hyun-Joo \surname{Kim}}\email{hjkim21@knue.ac.kr}
\author{Doil \surname{Jung}}
\affiliation{
$^1$ Department of Physics Education, Korea National University of Education,
Chungbuk 363-791, Korea\\
 }

\received{\today}

\begin{abstract} 

   Covariance matrices of heights measured relative to the average height of growing self-affine surfaces in
 the steady state are investigated in the framework of random matrix theory. We show that the spectral density 
 of the covariance matrix scales as $\rho(\lambda) \sim  \lambda^{-\nu}$ deviating from the prediction
 of random matrix theory and has a scaling form, 
 $\rho(\lambda, L) = \lambda^{-\nu} f(\lambda / L^{\phi})$ for the lateral system size $L$, 
 where the scaling function $f(x)$ approaches a constant for $\lambda \ll L^\phi$ and zero for 
 $L^\phi \ll \lambda < \lambda_{max}$. 
 The values of exponents obtained by numerical simulations are $\nu \approx 1.67$ and $\phi \approx 1.53$ 
 for the Edward-Wilkinson class and $\nu \approx 1.59$ and $\phi \approx 1.75$ for 
 the Kardar-Parisi-Zhang class, respectively. The distribution of the largest eigenvalues follows a scaling
 form as $\rho(\lambda_{max}, L) = 1/L^b f_{max} ((\lambda_{max} -L^a)/L^b)$, which is different from the Tracy-Widom
 distribution of random matrix theory while the exponents $a$ and $b$ are given by the same values for 
 the two different classes.
 
\end{abstract}

\maketitle

Over recent decades, growth phenomena of fluctuating interfaces persists as a fascinating subject of 
statistical physics. Fluctuating interfaces are among the most well studied non-equilibrium systems due to
their simplicity as well as ubiquity in nature and fundamental science \cite{growthreview}.
Growth of interfaces governed by local rules typically lead to the formation of self-affine
surfaces with universal scaling exponents for the surface width $W(L,t)$ which is defined as the standard 
deviation of interface height over a system size $L$. 
The surface width characterizes the roughness of the interface and follows a scaling
behavior $ W(L,t) = L^{\alpha}f(t/L^z)$, where the
scaling function $f(x)$ approaches to a constant for $x \gg 1$,  and $f(x) \sim  x^\beta$ for 
$x \ll 1$ with the dynamic exponent $z  = \alpha/\beta$  \cite{fv}. The exponents $\alpha$, and $\beta$ 
which are called the roughness and the growth exponent, respectively, which determine the universality 
classes of various 
fluctuating interfaces. The well-known universality class of the growing interfaces is the 
Kardar-Parisi-Zhang (KPZ) one which is predicted by the nonlinear Langevin equation \cite{kpz} due to
the slope dependent growth with the values of exponents $\alpha = 1/2$ and $\beta = 1/3$ for one dimension
and was widely confirmed in numerical models \cite{growthreview}.
While lacking of nonlinearity in the growth process results in another universality class called
the Edward-Wilkinson (EW) class where the values of exponents are given by $\alpha = 1/2$ and $\beta = 1/4$ 
for one dimension \cite{ew}.

The recent studies of fluctuating interfaces have dealt with other important characteristics 
beyond those for the scaling properties of the surface width. These include the distribution of the surface 
width \cite{widthdist}, and maximal and minimal height distributions \cite{heightdist,MRH}, etc. Especially, 
the asymptotic distribution of KPZ height fluctuations for curved initial conditions has been computed exactly 
for solvable models \cite{TW-KPZ} and found that it follows the Tracy-Widom
distribution (TWD) \cite{TW} of a Gaussian unitary ensemble, which has been confirmed by a recent experiment 
on the electro-convection \cite{TW-EXP} and the simulations \cite{TW-N}.
While for the flat initial condition, it is confirmed analytically
\cite{TW-flatA} and numerically \cite{TW-flatN} that the KPZ height distribution exhibits
the TWD of a Gaussian orthogonal ensemble. 
Although TWD has been first obtained in the statistics of the largest eigenvalues of random 
matrices belonging to the Gaussian ensembles \cite{TW}, it has been applied to other areas \cite{TWO} and its
application to the growth phenomena has given the connection between the growth problem and the random matrix 
theory (RMT) \cite{Mehta}. 

RMT was initially proposed to explain the statistical properties of nuclear spectra \cite{Mehta} but the 
usefulness of it in understanding the statistical properties of a complex system makes RMT be applied 
in the various systems \cite{Guhr,Forr}.
The statistical properties of matrices with independent random elements have been well described by the RMT and 
one can understand the statistical properties of a system by comparing the spectral statistics of  
the system with the results of RMT. The empirical cross-correlation matrices appearing in the 
study of various complex systems such as  the price fluctuations in the stock market \cite{stock}, electro encephalogram(EEG) data of brain \cite{eeg}, variation
of various atmospheric parameters \cite{atomo}, biophysical issues \cite{bio} and complex network \cite{network}, 
have been analyzed in the framework of RMT. 
 
The random Wishart matrix that is one of the standard tools in the RMT is defined via the product
$W = \frac{1}{N}X X^{\dagger}$ of a $M \times N$ random matrix $X$ having its elements drawn independently from a Gaussian distribution \cite{wishart}. If $X$ is a matrix whose elements represent some empirical data, 
then the Wishart matrix represents a empirical covariance matrix of the data, and the nondiagonal 
elements $W_{ij}$ of the covariance matrix have a direct interpretation as cross-correlation coefficients 
between data $X_i$ and $X_j$. Therefore if a certain complex system shows the spectral statistics same as
those of the Wishart matrix, one may think there are not significant correlations, while if it shows the
different properties from that, it could be regarded as there are some correlations.
However, the random and the correlated properties are too entangled in a real complex systems
to be simple elucidating the properties of correlations from the RMT analysis
of corresponding empirical covariance matrices. It informs just whether it is totally 
random or not. If the RMT is applied to the problem of fluctuating interfaces in which the strong 
correlation of the variables of heights
has been well understood would provide important insights into the universality of RMT for such a correlated system
as well as novel criteria for statistical properties of a fluctuating interface.

In this perspective, we would investigate further statistical properties of fluctuating interfaces under 
the framework of the RMT in this study. We construct the Wishart matrices by the product of the matrices 
having their elements of 
relative heights obtained from two models which belong to the KPZ and the EW universality classes, respectively.
The distributions of eigenvalues and the largest eigenvalues for each Wishart matrix are
 measured and compared
with the results of the RMT. The obtained results have shown the totally different properties of those of RMT and 
we have found new scaling features of the distributions.

Once the actual height $h_i(t)$ at the site $i$ and the time $t$ is generated, we define the relative 
height, $H_{it}=h_i (t) - \left<h_i (t )\right>$, where the spatially averaged height $\left<h_i (t )\right>$
keeps on growing with time and by subtracting it, the relative height has the zero mean and 
the distribution of relative height reaches a stationary state in the late-time regime in a finite system.
We consider the $L \times T$ height matrix $H$ with elements $H_{it}$, where $L$ is the lateral system size and $T$
is the time interval we considered.  We then compute the product 
symmetric matrix $C = \frac{1}{T} HH^\dagger$ with elements 
\be
C_{ij}  = \frac{1} {T} \sum_{t=1} ^T H_{it} H_{jt},
\ee
which represents the covariance matrix of heights and contains informations about height correlations
between two sites.

In the random matrix studies of eigenvalue spectra, the most popular property is the spectral density
or the distribution of eigenvalues $\rho (\lambda)$. For the Wishart random matrix \cite{wishart}, it was shown 
analytically that the spectral density $\rho(\lambda)$ is given by the Marcenko-Pastur(MP) 
law in the limit for $M, N \rightarrow \infty$ \cite{MP}:
 \be
 \rho_{MP}(\lambda) = \frac{\sqrt{(\lambda_+ - \lambda)(\lambda-\lambda_-)}}{2 \pi \sigma^2 m \lambda}
 \label{MP}
\ee
\be
 \lambda_{\pm} = \sigma^2 ( 1 \pm \sqrt{m})^2 
 \ee
%%%%%%%%%%%%%%%%%%%%%%%%%%%%%%%%%%%%%%%%%%%%%%%%%%%%%%%%%%%%%%%%%%%%%%%%%%
\begin{figure}[ht]
\includegraphics[width=8cm]{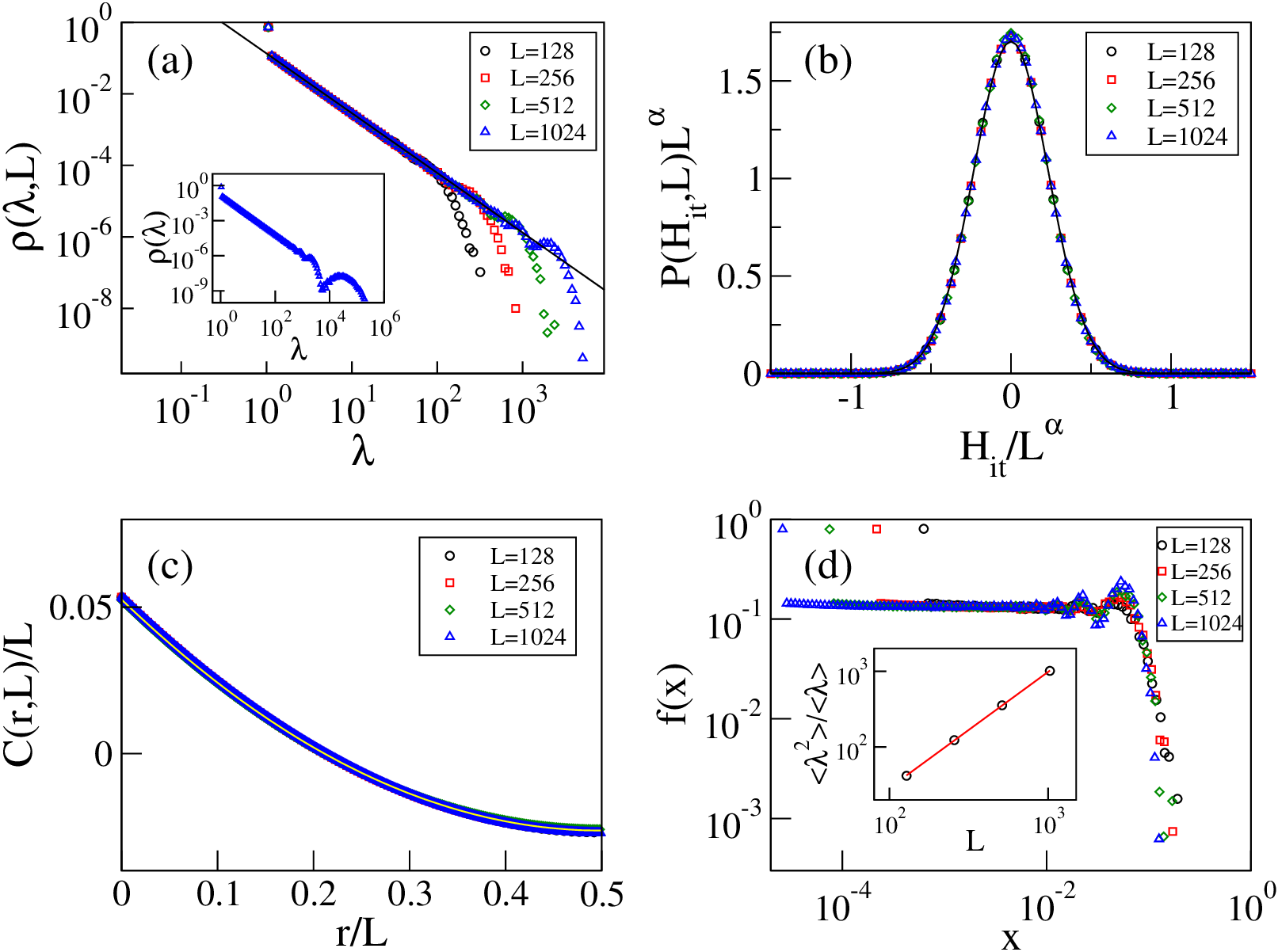}
\caption{(a) The inset shows the spectral density of eigenvalues $\rho(\lambda)$ of the covariance 
matrix of heights constructed by the EW model for $L=1024$. The plot is the spectral density
$\rho(\lambda, L)$ except for the largest eigenvalues for various $L$.
(b) The distributions of the elements of the height matrices $P(H_{it} , L)$ for the various $L$.
The solid line represents the Gaussian curve.
(c) The height correlations $C(r,L)$ for various $L$. The solid line represents the curve of Eq. (\ref{ceq})
(d) The plot shows the data collapse of $\rho(\lambda, L)$ and the plot
of $\left<\lambda^2 \right> / \left<\lambda \right>$ versus $L$ is shown in the inset.
}
\label{ed_ew}
\end{figure}
%%%%%%%%%%%%%%%%%%%%%%%%%%%%%%%%%%%%%%%%%%%%%%%%%%%%%%%%%%%%%%%%%%%%%%%%%%
where $\sigma$ is a standard deviation for elements of a $M \times N$ random matrix and $m=M/N$.
 The spectral density $\rho_{MP}(\lambda)$ shows the important features vanishing at both edges 
 of the MP sea and
 exhibiting a sharp maximum near minimum edge. For finite $M$ and $N$, the abrupt cut-off of 
 $\rho (\lambda)$ is replaced by a rapidly-decaying edge \cite{FMP}.

To compare the spectral properties of the covariance matrix of heights
with those of a random Wishart matrix, we have begun with the heights profiles generated by the 
EW model \cite{ew} which belongs to the EW universality class. 
 We have obtained the numerical data for the one dimensional substrate with the system size $L=128, 256, 512$, and
 $1024$. Here we have only focused on the properties in the late-time regime in which the spacial correlation 
 of heights is dominant and the correlation is independent on the time. 
 Thus we counted the time interval $T$ from the saturation time to the final time.
 
The inset of Fig. \ref{ed_ew} (a) shows the spectral density $\rho(\lambda, L)$ of the covariance matrix 
of heights constructed by the EW model for $L=1024$. It shows the totally different feature from MP law. 
The spectral density $\rho(\lambda)$ follows a power-law behavior except for large eigenvalues
most which are due to the largest eigenvalues.
So we considered the largest eigenvalue and the rest separately. Figure \ref{ed_ew} (a) shows the spectral
density excluded the largest eigenvalue for various system sizes.
The straight guide line represents that the spectral density obeys $\rho(\lambda) \sim \lambda^{-\nu}$ with 
$\nu = 1.67 \pm 0.02$. 
It is reasonable that the null hypothesis of no true correlation is rejected for the fluctuating interface
which is a strongly correlated system. In addition to,
 it shows the difference from the spectral density of the empirical covariance matrices most which 
 have just appeared the distortion of the shapes of the bulk of spectral density \cite{chaos,network}
 or abnormal largest eigenvalues \cite{stock}. The power-law behavior of the spectral density has been observed
 empirically in some systems such as the EEG data without the value of the exponent exactly measured\cite{eeg}. 
 Analytical argument for it was proposed with the L{\'e}vi matrices \cite{Levi, Levi-CB} where matrix elements 
 are distributed according to  $P(X_{ij})$ with $P(X_{ij}) \sim |X_{ij}|^{-(1+\mu)}$. For $\mu > 2$ 
 the distribution has finite variance while the variance diverges for $0 < \mu \leq 2$. In the case of the Wishart
 L{\'e}vi matrices, the distribution of eigenvalues has fat tails unlike the prediction
 of the MP law \cite{Wishart-Levi}. 
 In Ref. \cite{Levi-Burda}, the multivariate Student distribution has been used as the 
 power-like distribution and it has been obtained that the spectral density $\rho(\lambda)$ of
 the Wishart L{\'e}vi matrices decays like $\lambda ^{-(\mu /2 +1)}$.

To check whether our spectral density follows the scheme of Levi matrix, we have measured the 
distributions of the elements of the height matrix $P(H_{it},L)$ for the various $L$. 
As shown in Fig. \ref{ed_ew} (b)
the distributions $P(H_{it}, L)$ has a scaling form, 
$P(H_{it}, L) = 1/L^\alpha g\left(H_{it}/L^\alpha \right)$ and fall on a single curve, 
where $g(x)$ is the Gaussian curve and $\alpha = 0.5$. 
It results from that the only relevant scale is the roughness $W$ 
being consistent with the scaling description in the late-time regime \cite{MRH,hc}.
Thus $P(H_{it})$ has no power-law tails,
which means that the origin of the power-law behavior of $\rho(\lambda)$
is not in the fat tail of the distribution of matrix elements. 
And it shows that the MP law might be not valid any more even when the distribution of matrix 
obeys the Gaussian distribution.

The elements of a L{\'e}vi matrix are uncorrelated  
random variables, while in the late-time regime, the elements of the height matrix, {\it i.e.}, 
the relative heights at different sites are strongly correlated as a following equation \cite{hc} 
\be
C(r,L)=\left< H_{it}H_{(i+r)t} \right> \sim L \left[1-\frac{6r}{L}(1-\frac{r}{L})\right].
\label{ceq}
\ee
Figure \ref{ed_ew} (c) shows the height correlations $C(r,L)$ measured by averaging the 
elements of the covariance matrix $C_{ij}$ having $r = |i-j|$ which are excellent agreement
with Eq. (\ref{ceq}) for various system size $L$. 
It indicates that the covariance matrix of heights has elements decreasing away from the
main diagonal like Eq. (\ref{ceq}). The long-ranged correlation of the elements
might give rise to the power-law behavior of eigenvalue density $\rho(\lambda)$.
It is comparable to the power-law random banded matrix (PRBM) model \cite{CrRME} which is defined
as the ensemble of matrices with elements
\be
M_{ij} = G_{ij} a(|i-j|),
\ee
where the matrix $G$ runs over the GOE and $a(r) \sim r^{-\alpha}$ for large $r$.
This exhibits an Anderson localization transitions at $\alpha=1$ and allowed a detailed study 
of the wave function and energy-level statistics at criticality.
Although the PRBM model represents ensembles of matrices with long-ranged 
off-diagonal disorder, its spectral properties are different from those of the covariance matrix of
heights and thus it may serve as a another new critical random-matrix ensemble
with long-ranged off-diagonal random hopping like Eq. (\ref{ceq}).  
   
%%%%%%%%%%%%%%%%%%%%%%%%%%%%%%%%%%%%%%%%%%%%%%%%%%%%%%%%%%%%%%%%%%%%%%%%%%
\begin{figure}[ht]
\includegraphics[width=8cm]{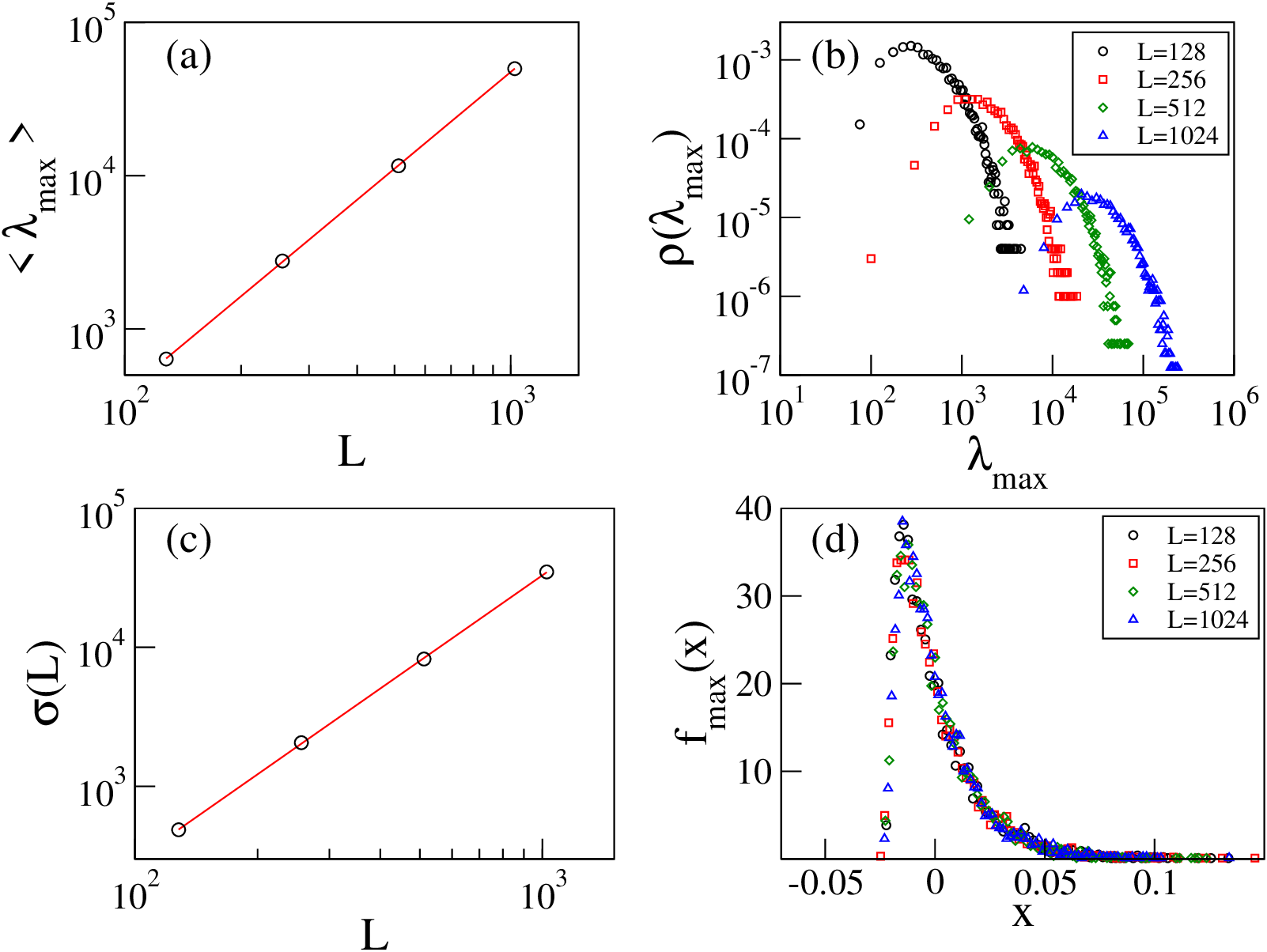}
\caption{(a) The average of the largest eigenvalues $\left<\lambda_{max} \right> $
as a function of the system size $L$ for the EW model.
(b) The distribution of the largest eigenvalues $\rho(\lambda_{max})$.
(c) The standard deviation $\sigma (L)$ of $\lambda_{max}$.
(d) The data collapse of $\rho(\lambda_{max},L)$ for the various $L$.
}
\label{em_ew}
\end{figure}
%%%%%%%%%%%%%%%%%%%%%%%%%%%%%%%%%%%%%%%%%%%%%%%%%%%%%%%%%%%%%%%%%%%%%%%%%%

On the other hand, we found that the finite-size distribution of eigenvalues $\rho(\lambda, L)$
for the various system size $L$ obeys a scaling form of the type,
\be
\rho(\lambda, L) = \lambda^{-\nu}f\left(\frac{\lambda}{\lambda_c (L)}\right)
\label{sf}
\ee
where $\lambda_c (L)$ is a characteristic eigenvalue which scales as $\lambda_c (L) \sim L^\phi$.
$f(x)$ is a scaling function satisfying the following properties:
\be
\begin{aligned}
f(x) & = \text{const},   & \text{for }   & \lambda \ll L^\phi   \\
f(x)  &\rightarrow 0,    &\text{for }    & L^\phi \ll \lambda < \lambda_{max}.
\end{aligned}
\ee
Figure \ref{ed_ew} (d) shows the data collapse of the spectral density 
$\rho(\lambda, L)$ with $\nu = 1.67$ and
$\phi = 1.53$ and they fall on a single curve very well except for small eigenvalues. 

The exponent $\phi$ can be measured by an alternative way.
The n-th moment of the eigenvalue is obtained by
\be
\begin{aligned}
\left<\lambda^n \right> &= \int_0 ^{\lambda_2} \lambda^n \rho(\lambda, L) d\lambda \\
&= L^{\phi (n+1-\nu)} \int_0 ^{x_2} x f(x) dx
\end{aligned}
\ee
where $\lambda_2$ is the second largest eigenvalues and $x_2 = \lambda_2 /L^\phi$.
The integral has a finite value and we obtain
\be 
\left<\lambda^n \right> \sim L^{\phi(n+1-\nu)}.
\ee
Thus we can provide that
\be
\frac{\left<\lambda^2 \right>}{\left<\lambda \right>} \sim L^\phi.
\ee
We measured $\phi = 1.53\pm 0.03$ from it as shown in the inset of Fig. \ref{ed_ew} (d) which
is good agreement with the value used in the collapse of $\rho(\lambda, L)$.
 
Next, we compared the properties of the largest eigenvalues of the covariance matrix of heights with
the results of the RMT. 
The MP law tells that the average of the largest eigenvalue $\left<\lambda_{max} \right> $ depends on the matrix
size $M$ like as $\left<\lambda_{max} \right> \sim M$ for large $M$ and the typical fluctuations of 
$\lambda_{max}$ are known to be described by the TWD \cite{TW,Majundar}.
We measured the average of largest eigenvalue $\lambda_{max} (L)$ as a function of $L$ and found that
it scales as $\left<\lambda_{max}\right> \sim L^{a}$ with $a \approx 2.09$ (Fig. \ref{em_ew} (a)), 
which is also different from the result of the RMT.
Also the distributions of the largest eigenvalue $\rho(\lambda_{max})$ 
deviates from the TW curve as shown in the Fig. \ref{em_ew} (b). 
Hence it does not follow the prediction of the RMT.

The fluctuations of $\lambda_{max}$ from its the average value come to be larger as $L$ increases. 
We measured the standard deviation $\sigma (L)$ of $\lambda_{max}$ and found that 
$\sigma (L)$ scales as $\sigma (L) \sim L^b$ with $b \approx 2.04$. 
We rescaled the variable $\lambda_{max}$ as $x = (\lambda_{max} - L^a)/{L^b}$ 
and obtained a scaling form of $\rho(\lambda_{max}, L)$ as follows, 
\be
\rho(\lambda_{max}, L) = \frac{1}{L^b} f_{max}(\frac{\lambda_{max} - L^{a}}{L^{b}}).
\label{fmax}
\ee
Figure \ref{em_ew} (d) shows that the distributions of largest eigenvalues $\rho(\lambda_{max}, L)$
fall on a single curve with $a=2.09$ and $b=2.09$ for the different system size $L$.
Thus, the analysis of the fluctuating interfaces belonging to the EW universality class 
under the framework of the RMT
shows the results different from the conventional behaviors of the RMT 
and the new spectral scaling properties for the EW class. 
%%%%%%%%%%%%%%%%%%%%%%%%%%%%%%%%%%%%%%%%%%%%%%%%%%%%%%%%%%%%%%%%%%%%%%%%%%
\begin{figure}[ht]
\includegraphics[width=8cm]{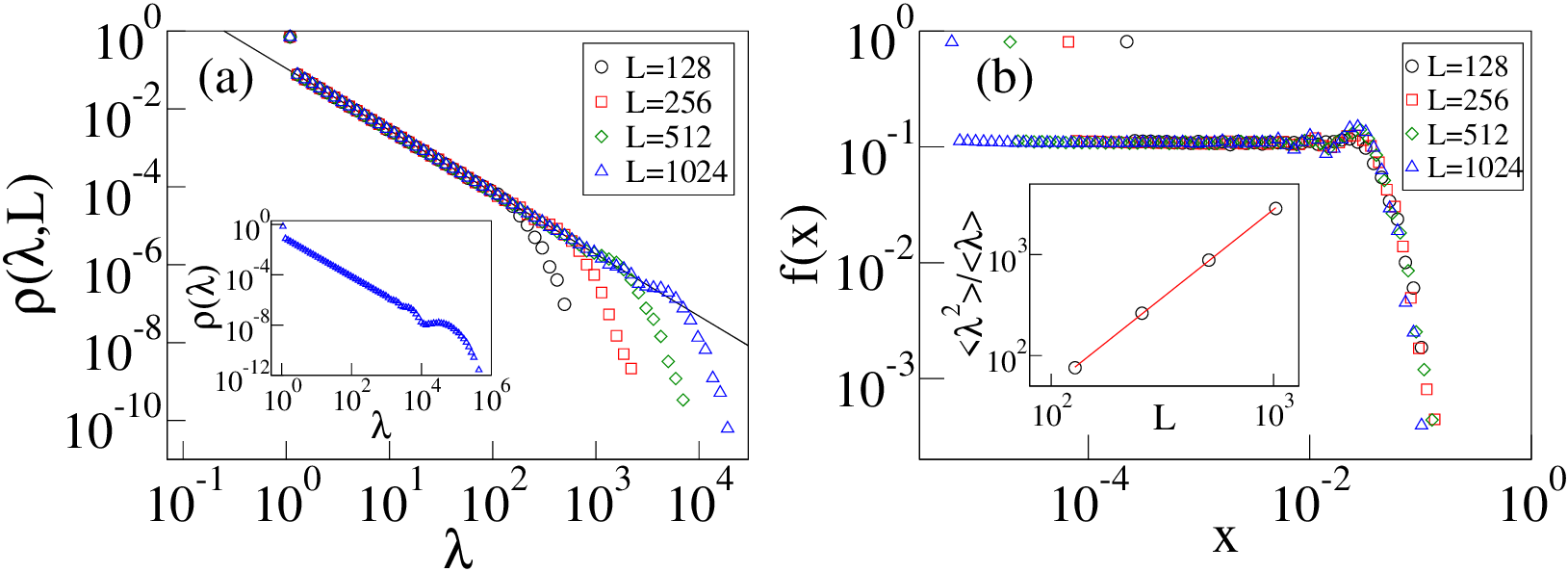}
\caption{(a) The spectral density $\rho(\lambda, L)$ of the covariance matrix of heights 
constructed by the RSOS model except for the largest eigenvalue. 
The inset shows $\rho(\lambda)$ including the largest eigenvalue for $L=1024$.  
(b) The data collapse of $\rho(\lambda, L)$ for various $L$.
The inset shows the plot of $\left<\lambda^2 \right> / \left<\lambda \right>$ versus $L$.
}
\label{ed_rsos}
\end{figure}
%%%%%%%%%%%%%%%%%%%%%%%%%%%%%%%%%%%%%%%%%%%%%%%%%%%%%%%%%%%%%%%%%%%%%%%%%%

By applying the RMT to the analysis of the another universality class of the growth 
phenomena, the KPZ class, we would like to compare the spectral properties
between two universality classes.
We constructed the covariance matrix of the heights generated by the restricted-solid-on-solid (RSOS) 
model \cite{rsos} which belongs to the KPZ universality class. 
The spectral density of this covariance matrix $\rho(\lambda)$ also follows a power-law behavior like
the case of the EW model (Fig. \ref{ed_rsos} (a)).
%%%%%%%%%%%%%%%%%%%%%%%%%%%%%%%%%%%%%%%%%%%%%%%%%%%%%%%%%%%%%%%%%%%%%%%%%%
\begin{figure}[ht]
\includegraphics[width=8cm]{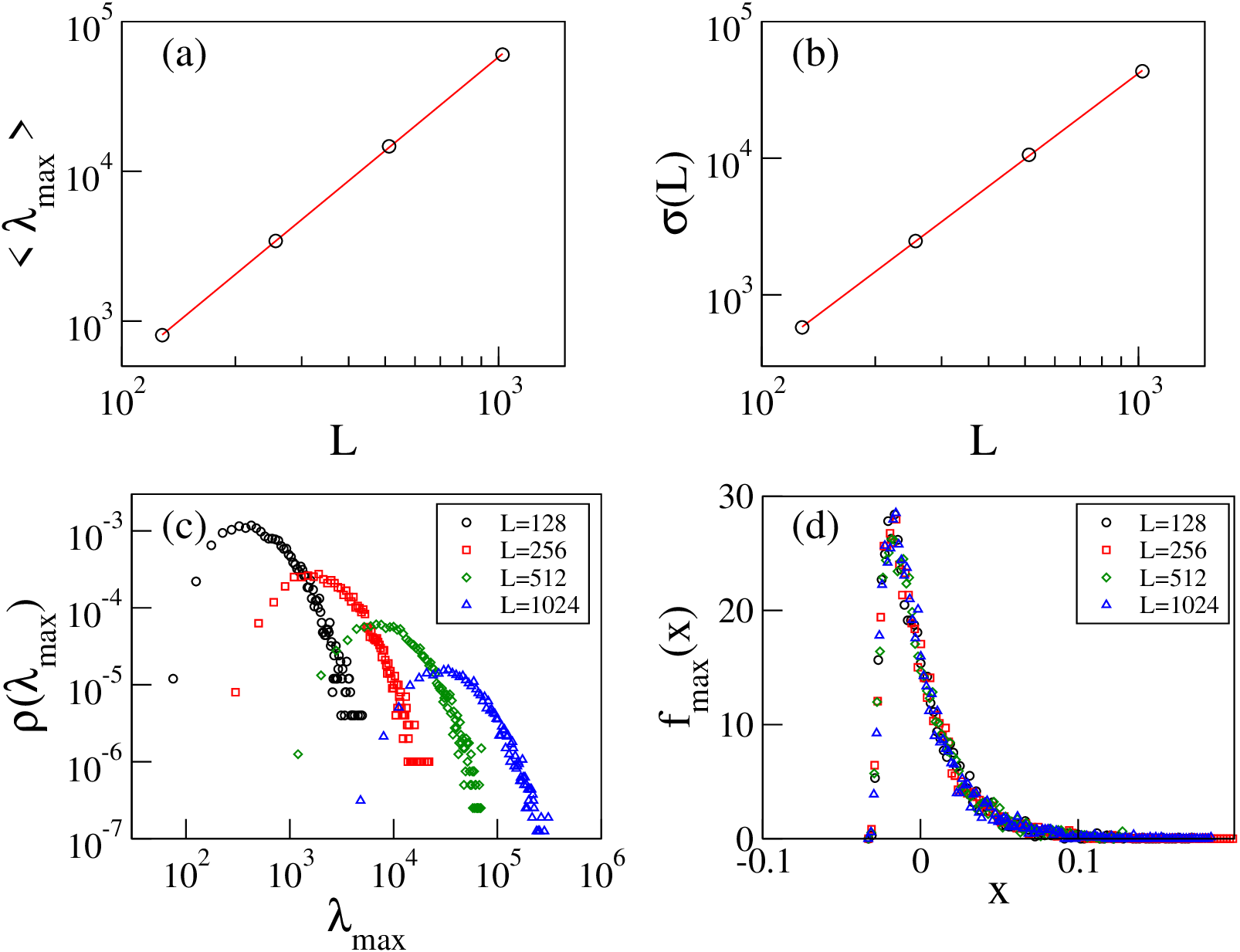}
\caption{ (a) The average of the largest eigenvalues $\left<\lambda_{max} \right> $
as a function of the system size $L$ for the RSOS model.
(b) The standard deviation $\sigma (L)$ of $\lambda_{max}$.
(c) The distribution of the largest eigenvalues $\rho(\lambda_{max})$.
(d) The data collapse of $\rho(\lambda_{max},L)$ for various $L$.}
\label{em_rsos}
\end{figure}
%%%%%%%%%%%%%%%%%%%%%%%%%%%%%%%%%%%%%%%%%%%%%%%%%%%%%%%%%%%%%%%%%%%%%%%%%%
However the value of exponent $\nu$ was obtained
as $\nu = 1.59 \pm 0.01$, which is different from that of the EW model. The height correlation of the
RSOS model also obeys the Eq. (\ref{ceq}) which indicates that the value of $\nu$
depends on the details besides the height correlations.
The spectral density $\rho(\lambda,L)$ for various $L$ also follows the scaling form of Eq. (\ref{sf}).
The data collapse of the spectral density $\rho(\lambda,L)$ for the RSOS model 
is shown in the Fig. \ref{ed_rsos} (b). 
The values of the exponent $\nu$ and $\phi$ were taken by 1.59 and 1.75, respectively.
The inset of Fig. \ref{ed_rsos} (b) shows the plot of $\left<\lambda^2 \right> / \left<\lambda \right>$
versus $L$ and we obtained $\phi = 1.75 \pm 0.03$ of which value is also different from that of the EW model.
 
The properties of the largest eigenvalues for the RSOS model are shown to be similar to those for
the EW model. The average of the largest eigenvalues scales as $\left<\lambda_{max}\right> \sim L^{a}$ with 
$a \approx 2.08$ (Fig. \ref{em_rsos} (a)) and the standard deviation $\sigma (L)$ of $\lambda_{max}$ 
scales as $\sigma (L) \sim L^b$ with $b \approx 2.08$ (Fig. \ref{em_rsos} (b)). 
Figure \ref{em_rsos} (c) shows the distribution of the largest eigenvalues which does not follow the
TW curve like the EW model for the various system size $L$. 
The data collapse of $\rho(\lambda_{max}, L)$ by the scaling form of Eq. (\ref{fmax}) falls on a
single curve with $a=2.08$ and $b=2.08$ (Fig. \ref{em_rsos} (d)).

In summary, we investigated the spectral properties of covariance matrices of relative heights of
fluctuating interfaces in the late-time regime. 
The spectral density of the covariance matrices follows the power-law behavior except for the 
largest eigenvalue, which is different from the MP law of RMT. It indicates that the random variables 
correlated like the relative height of a fluctuating interface would give rise to 
the power-law behavior of the spectral density of the corresponding covariance matrix. 
As a finite-size effect the spectral density falls zero beyond the characteristic eigenvalue
depending on the lateral system size and has the scaling form. The values of exponents $\nu$ and $\phi$
related to the scaling form were given by the different values for the EW model and the RSOS model.
It indicates that the statistical characteristics of the different classes of two models 
in the late-time regime are reflected to the specral properties of them.
The distribution of the largest eigenvalues also showed the different features from the TWD. 
The average and the standard deviation of the largest eigenvalues follow the power-law behaviors
with the lateral system size. The values of the scaling exponents do not give the difference
between the EW model and the RSOS model. 
It would be desirable if it is applied to the behavior in the short-time regime
and various interface models belonging to the different universality classes
and the local spectral properties of the critical random matrix ensembles 
using the correlated heights are further studied.

This work was supported by the 2012 research grant from Korea National University of Education.

\end{document}